\documentclass[%
reprint,
nofootinbib,
 amsmath,amssymb,
 aps,
prstab,
floatfix,
]{revtex4-2}

\usepackage{graphicx}
\usepackage{dcolumn}
\usepackage{bm}
\usepackage{graphicx}
\usepackage{dcolumn}
\usepackage{bm}
\usepackage[caption=false]{subfig}
\DeclareCaptionJustification{justified}{\justifying}
\usepackage[dvipsnames]{xcolor}
\usepackage{comment}
\usepackage{hyperref}
\DeclareUnicodeCharacter{00A0}{~}

\newcommand{\AT}[1]{{\color{ForestGreen}#1}}
\newcommand{\DS}[1]{{\color{Cyan}#1}}
\begin{document}

\preprint{APS/123-QED}

\title{An Innovative Transverse Emittance Cooling Technique using a Laser-Plasma Wiggler}
	
\author{$^{1,\dagger}$O.~Apsimon}
\thanks{email: oznur.apsimon@liverpool.ac.uk}
\author{$^2$D.~Seipt}
\author{$^{1,5,\dagger}$M. Yadav}
\author{$^{1,\dagger}$A. Perera}
\author{$^3$Y.~Ma}
\author{$^{4,\dagger}$D.A.~Jaroszynski}
\author{$^3$A.~G.~R.~Thomas}
\author{$^{6,\dagger}$G.~Xia}
\author{$^{1\dagger}$C.~P.~Welsch}
\affiliation{$^1$The University of Liverpool, Liverpool L69 3BX, United Kingdom}
\affiliation{$^2$Helmholtz-Institut Jena, Fröbelstieg 3, 07743 Jena, Germany}
\affiliation{$^3$Gérard Mourou Center for Ultrafast Optical Sciences and Department of Nuclear Engineering and Radiological Sciences, University of Michigan, Ann Arbor, MI 48109, USA}
\affiliation{$^4$University of Strathclyde, Glasgow G11XQ, Scotland}
\affiliation{$^5$University of California, Los Angeles, CA 90095, USA
}
\affiliation{$^6$ University of Manchester, M13 9PL, Manchester, United Kingdom}
\affiliation{$^{\dagger}$ The Cockcroft Institute of Accelerator Science and Technology, Warrington, WA4 4AD, United Kingdom}

\begin{abstract}
We propose an innovative beam cooling scheme based on laser driven plasma wakefields to address the challenge of high luminosity generation for a future linear collider. For linear colliders, beam cooling is realised by means of damping rings equipped with wiggler magnets and accelerating cavities. This scheme ensures systematic reduction of phase space volume through synchrotron radiation emission whilst compensating for longitudinal momentum loss via an accelerating cavity. In this paper, the concept of a plasma wiggler and its effective model analogous to a magnetic wiggler are introduced; relation of plasma wiggler characteristics with damping properties are demonstrated; underpinning particle-in-cell simulations for laser propagation optimisation are presented. The oscillation of transverse wakefields and resulting sinusoidal probe beam trajectory are numerically demonstrated. The formation of an order of magnitude larger effective wiggler field compared to conventional wigglers is successfully illustrated. Potential damping ring designs on the basis of this novel plasma-based technology are presented and performance in terms of damping times and footprint was compared to an existing conventional damping ring design. 
\end{abstract}

\maketitle



\section{Introduction} 
After the discovery of the Higgs boson at the Large Hadron Collider (LHC) at CERN \cite{ATLAS_Higgs,CMS_Higgs}, the global vision is a successor collider in order to explore the properties of this newly discovered boson in great detail. The physics of the Higgs field still holds secrets of fundamental interactions and can only be addressed in the controlled environment of an electron-positron collider that allows elementary particle collisions with minimum synchrotron radiation losses and well defined initial momenta.  Over the last few decades, two different design approaches, using normal conducting (CLIC, \cite{CLIC}) and superconducting (ILC, \cite{ILC}) radio frequency structures, revealed the size of such a machine to be about 30$-$50$\,$km. The electron-electron option of the Future Circular Collider (FCC) project envisages a 100\,km circumference machine for collisions at centre-of-mass energy of 90\,GeV$-$350\,TeV \cite{FCC}. The limits of conventional technologies for high energy demand of particle physics exploration persuade us to new frontiers of particle accelerator science. Paradigm shifting technologies are being developed such as plasma acceleration \cite{8GeV, awake, energy_doubling}. However, there are still many challenges to be addressed before the maturation of plasma technology for large-scale accelerator applications. Advanced and novel accelerators community (ICFA- ANAR2017) underlined and prioritised the following technological challenges; the repetition rate, efficiency and beam quality of the lasers; scalability of the system; delivery of high collision luminosity; resolution of diagnostics; comprehensive simulations and the availability of dedicated test facilities \cite{ANAR19}. Driven by high-power lasers, plasma can generate orders of magnitudes larger electric fields compared to metallic radio-frequency cavities \cite{8GeV, Esarey, Hooker}. In this paper, an innovative cooling method is presented which shows great promise of replacing the magnetic wigglers in a damping ring with plasma wigglers and providing superior beam quality for the future linear collider based on plasma technology.

In order to achieve high luminosity, a the ILC design relies on emittance cooling via damping rings followed by a long transfer line from ring to main linear accelerator followed by beam delivery system to the final interaction point. Emittance cooling is achieved through radiation damping due to synchrotron radiation emitted from beam particles moving along curved trajectories of a circular accelerator (damping ring) that has a circumference of a few hundred meters to kilometers. Particles emit synchrotron radiation depending on the local curvature of the orbit within a cone of angle $\/\gamma$, where $\gamma$ is the relativistic Lorentz factor. Longitudinal momentum is restored by radio-frequency cavities in the ring, while transverse momentum is damped on every turn through radiation damping and quantum excitation are equal and an equilibrium value is reached \cite{handbook}, typically in milliseconds \cite{ILC_damping_time, sys_damp_design}. A faster damping is achieved by increasing the energy loss per turn by adding high field periodic magnetic structures (wigglers or wigglers, depending on the magnetic strength) \cite{wiggler_damping}. 

As an example, the International Linear Collider (ILC) project proposes a machine with 250-500$\,$GeV centre-of-mass energy over about a 31$\,$km footprint \cite{ILC_tech_des_report}. Electrons and positrons emerging from different sources undergo an initial acceleration up to 5$\,$GeV before they are injected into a their respective damping rings with a circumference of 3.2$\,$km, housed in the same tunnel. ILC damping rings are designed in a race track shape to accommodate two straight sections. A radiative section comprising 54 super-ferric wigglers is located in one of these straight sections. Each wiggler is 2.1$\,$m long and generates a 2.16$\,$T peak magnetic field when operating at 4.5$\,$K and radiates 17 kW radiation power \cite{ILC_tech_des_report}. This straight section also houses a superconducting radio-frequency system to replenish the longitudinal momentum of the beam. 


A compact synchrotron radiation insertion device with significantly larger fields than the current wigglers would limit the footprint and the cost of the ring by reducing the required number and size of these insertion devices. There are several methods for plasma assisted radiation generation from a relativistic particle beam, such as, betatron oscillations, Compton scattering, bremsstrahlung and transition radiation \cite{Alec2016}. We propose to incorporate the concept of plasma wiggler as radiators in a damping ring to benefit from their large effective magnetic fields and compactness. There are various concepts to conceive a plasma wiggler \cite{PRL114,plasma_undulator1,plasma_undulator2,plasma_undulator3}. Following the one proposed in \cite{PRL114}  , a plasma wiggler is formed when a short laser pulse is injected into plasma off-axis or at an angle that causes the centroid of the laser pulse to oscillate. Given that the product of the plasma wave number and the characteristic Rayleigh length of the laser is much larger than one, the ponderomotively driven plasma wake will follow this centroid. This oscillating transverse wakefield works as an wiggler forcing particles to follow sinusoidal trajectories and emit synchrotron radiation. In addition, the damping time is inversely proportional to the square of the magnetic field of the damping device. Theoretically, a plasma wiggler can generate order of magnitude larger effective magnetic fields than conventional wigglers, hence can reduce the length of the damping units by a factor of hundred while providing the same damping times. 

In this paper, we layout the fundamental design criteria including discussion on laser-plasma channel coupling and average dephasing length in parabolic channel to achieve a plasma wiggler in Section \ref{optimisation}. This is followed by numerical demonstration of oscillation laser centroid in the channel and sinusoidal probe beam trajectory in Section \ref{numerical}. In this section, we present the baseline design that achieves 20\,T effective magnetic field compared to a typical ~2\,T in a superconducting magnetic wiggler, as well as 4\,MW average radiated power by two orders of magnitude improvement from the conventional approach. This also includes studies discussing correlation of laser injection offset with laser radius and plasma skin depth laying out interesting phenomena of radial phase shift in focusing force. In Section \ref{damping}, we presented three different damping ring designs that offer compromises between low damping time or small footprint. 

\section{Plasma wiggler}
\label{optimisation}
The optimisation criteria for a plasma wiggler include laser and plasma parametrisation ensuring the matching between the laser spot size and channel radius; probe beam emittance to the channel width as well as limitations such as dephasing length in a channel. In this section, characteristics of a plasma wiggler following the optimisation criteria is numerically demonstrated.

Laser propagation through preformed plasma channels were extensively studied, previously \cite{guiding1, guiding2,Dino1, Dino2, Dino3, Dino4, Dino5, Dino6, Dino7, Dino8}. A plasma wiggler is formed when a short laser pulse is injected into a parabolic plasma channel off-axis or at an angle that causes the centroid of this laser pulse to oscillate \cite{PRL114} when the normalised vector potential $a_0<1$. A Gaussian laser pulse will be guided in a parabolic channel in Eq.\ref{eqn:parabolic}, 
\begin{equation}
n(r) = n_0(1+(\Delta n/n_0)r^2/w_0^2),
\label{eqn:parabolic}
\end{equation}
at the matched spot size when the channel depth is equal to the critical channel depth as in $\Delta n =(\pi r_e w_0^2)^{-1}$ \cite{Esarey, guiding3}. In this expression $r_e=e^2/mc^2$ is the classical electron radius, $n_0$ is the on-axis plasma density and $w_0$ is the laser radius. 

For a laser with sufficiently low power to satisfy $P < P_c$, the self-focusing is avoided and the laser spot size is preserved during the propagation for an initial spot size matched to the channel radius. If the laser power is higher than the critical power, $P_c[GW]\approx17(k_L/k_p)^2$, then relativistic self-focusing will occur \cite{Esarey}, where $k_L$ and $k_p$ are the wave numbers for the laser and the plasma wave, respectively.
 
The period of the laser centroid oscillation is determined by the Rayleigh length of the laser, $Z_R=\pi w_0^2/\lambda_L$, and given by $2\pi Z_R$. According to this, the centroid oscillation follows,
\begin{equation}
x = x_{ci}\cos(z/Z_R + \phi),
\end{equation}
where $\phi$ is the initial arbitrary phase and $x_{ci}$ is the initial transverse offset. The wiggler period, $\lambda_u$, is equal to the wavelength of the laser centroid oscillation,  
\begin{equation}
\lambda_u = 2\pi^2 w_0^2 / \lambda_L,
\label{eqn:lambdau}
\end{equation}
where $\lambda_L$ is the wavelength of the driver laser. 
 
Furthermore, when $k_pZ_R \gg 1$, the ponderomotively driven wakefields follow the oscillating laser centroid. In the case of an electron bunch injected into these oscillating wakefields at an appropriate phase such that the longitudinal field $E_x = 0$ and the transverse field $E_{y}$ is nonzero, then these probe electrons will follow oscillatory trajectories correlated with the laser centroid.  

The strength parameter corresponding to a plasma wiggler is derived in \cite{PRL114} and given in Eq. \ref{eqn:au}.
\begin{equation}
a_u \approx 4\pi a_0^2 C x_{ci} / \lambda_L
\label{eqn:au}
\end{equation}
where $C = \sqrt{\pi/8 e} \approx 0.38$ for an optimised laser pulse duration and $C \rightarrow 2C$ for a circularly polarised laser.

In a magnetic wiggler, assuming that the magnetic field is sinusoidal, integrating the equation of motion of the electrons in transverse plane, $\ddot{z}=eB_y/p$, yields the maximum horizontal deflection or the wiggler strength parameter,
\begin{equation}
K = \frac{B_0}{m_ec} \frac{e\lambda_u}{2\pi},
\label{eqn:K}
\end{equation}
where $B_y$ is the transverse field with a  peak value of $B_0$, $p$ is the momentum and $m_e$ is the mass of an electron.

Therefore, an effective magnetic field for a plasma wiggler can be defined in practical units as in 
\begin{equation}
B_{\textrm{eff}}[\textrm{T}] = a_u / (0.0934 \lambda_u[\textrm{mm}])
\label{eqn:beff}
\end{equation}
where $K$ is replaced by the strength of plasma wiggler, $a_u$ and $\lambda_u$ now representing the period length of the plasma wiggler. Eq. \ref{eqn:beff} can be translated into $B_{\textrm{eff}}[T]=a_0^2Cx_{ci}[m]/(46.7 \pi (w_0[m])^2 )$  by substituting Eq. \ref{eqn:lambdau} and Eq. \ref{eqn:au} in Eq. \ref{eqn:beff}. Therefore, theoretically, for a normalised vector potential, $a_0=0.36$, and the laser spot size, $w_0=10\mu$m, a plasma wiggler can generate an order of magnitude larger wiggler field strength than superconducting wigglers designed for ILC. For $a_0$ values close to unity, strong transverse wakefields might cause ejection of probe electrons outside the wiggler and act as a deflecting magnet as shown in Section \ref{numerical}. Therefore, higher effective wiggler fields may be achieved by reducing the laser spot size. However, one should note that for small values of $w_0$, fields created behind the laser pulse occupy a narrow transverse region. Therefore capturing probe electrons can be challenging. The general rule of thumb is that the transverse extend of both probe electrons and laser centroid offset should be smaller than the laser spot size.
 
\subsection{Coupling into the Channel} 
\paragraph{Laser Matching}
For the laser to propagate through the plasma with a constant spot size, the initial spot size and the channel radius should match. A detailed discussion is given in \cite{PRAB19} by introducing dimensionless variables in units of plasma frequency in SI units, $\omega_p=\sqrt{n_0 e^2 / m_e \epsilon_0}$ and the wave number, $k_p=\omega_p/c$. According to this, the evolution of the laser spot size, $w$, is given by, 

\begin{equation}
w^2 = \frac{w_0^2}{2} + \frac{2R^2}{w_0^2} + \left( \frac{w_0^2}{2} -  \frac{2R^2}{w_0^2} \right)\cos2\Omega\tau,
\label{eqn:laserspot}
\end{equation}
where $w_0$ is the laser spot size, $R$ is the dimensionless channel radius, $\Omega$ is the characteristic frequency of the laser centroid oscillation and $\tau$ is dimensionless time. The matched case, where the spot size, $w$, is constant, requires the second term to be zero. Hence, for a matched case $R=r_m^2/2$ where $r_m$ is the dimensionless matched laser radius.  Furthermore, $\Omega_m = 2/(M_p r_m^2)$ is the dimensionless frequency of the centroid oscillation, where $M_p = k_L/k_p$. The coefficient $M_p/2\pi$ transforms the space and time into units measured by plasma wavelength and period, respectively. 

\begin{figure}[htb!] 
\centering
\captionsetup{justification=justified}
\includegraphics[width=0.5\textwidth] {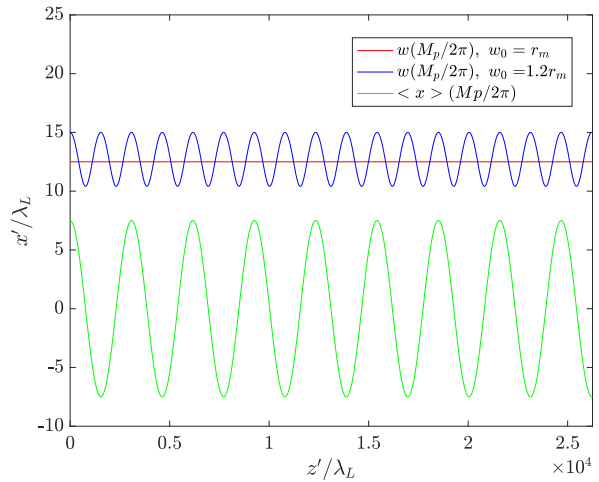}  
\caption{Evolution of the dimensionless laser spot size for the cases where the plasma channel radius is matched ($w_0=r_m$) and unmatched  ($w_0=1.2r_m$) to the laser spot size (red and blue curves, respectively). Laser centroid oscillations in the units of laser wavelength for the matched case is also presented with the green curve. The plot is generated for $a_0=0.36$, $w_0=10\,\mu$m and $n_0=7\times 10^{23}$ m$^{-3}$.}
\label{fig:matching}
\end{figure}
 
Figure \ref{fig:matching} presents the evolution of the laser spot size according to Eq.\ref{eqn:laserspot}, for matched ($w_0 = r_m$) and unmatched ($w_0 = 1.2r_m$) cases in units of laser wavelength. For the unmatched case (blue curve), laser spot size oscillates with an amplitude of $\pm$5$\lambda_L$. Whereas it is constant at 12.5$\lambda_L$ during its propagation for a matched condition. Figure \ref{fig:matching} also shows the evolution of the centroid for the matched case which oscillates at $\Omega_m$. 
 
 \paragraph{Beam Envelope Matching} The beam envelope evolves as a function of the beam emittance and energy as well as any external focusing or defocusing (such as space charge) and dispersive effects. Ignoring the adiabatic damping, any dispersive contribution and space charge, the beam size, $r_b$, is given as below. 
\begin{equation}
\frac{d^2r_b}{d(ct)^2} = \frac{\varepsilon_n^2}{\gamma^2 r_b^3} - k_{\beta}^2r_b
\label{eqn:envelope}
\end{equation} 
where $\varepsilon_n = \gamma r_b \theta_b$ is the normalised transverse beam emittance with $\theta_b$ is the rms beam angle and $\gamma$ is the Lorentz factor. To maintain a constant beam size through the channel, i.e.,  $d^2r_b/dct^2 =0$, the matched beam size should be $r_{bm} = (\varepsilon_n/\gamma k_{\beta})^{1/2}$ \cite{emitt_matching}. This matching condition is also employed for the simulations presented here.

\subsection{Dephasing Length in a Parabolic Channel}
Effective length of propagation for laser-plasma interactions is generally limited by laser diffraction. Guiding high power laser pulses through a parabolic plasma channel can circumvent this. Here, the channel acts as effective radio-frequency cavities that confine and shape the electromagnetic fields in conventional accelerators. When diffraction is eliminated, the limiting factor for the effective interaction length is the dephasing length,
\begin{equation}
\lambda_d = \lambda_p^3/\lambda_L^2,
\label{eqn:lambda_d1}
\end{equation}
where  $\lambda_p=2\pi/k_p$ is the on-axis plasma wavelength. This is the distance over which a trailing electron bunch gains sufficient energy to outrun the driving laser pulse to cross into decelerating fields. The dephasing length scales with $n_0^{-3/2}$ where $n_0$ is the plasma density on the channel axis. To prevent witness electrons overtaking the driver laser pulse, the dephasing length should be larger than, or similar to the length of the plasma wiggler. Therefore, one must optimise the wiggler target by establishing a compromise between the interaction length and plasma density \cite{Leemans98, Geddes04}. 

 However, in the case of a sinusoidal trajectory of a laser pulse in a channel, plasma density hence the dephasing length for the laser has a radial dependence. Therefore, such a laser pulse would experience an average dephasing length that is shorter than a laser travelling on channel axis through the channel. The axial dephasing length can be expanded as in Eq. \ref{eqn:lambda_d2}, 
\begin{equation}
\lambda_d(r) = \frac{1}{\lambda_L^2}(2\pi c)^3\bigg(\frac{\epsilon_0 m_e}{e^2}\bigg)^{3/2}n(r)^{-3/2},
\label{eqn:lambda_d2}
\end{equation}
by substituting the plasma wavelength and isolating the plasma density as a function of radial coordinate.
The average dephasing length is then calculated as,
\begin{equation}
 \bar{\lambda}=\frac{1}{y_i - y_f} \int^{y_f}_{y_i} \lambda_d(y)dy,
 \label{eqn:lambda_d3}
\end{equation}
resulting in,
\begin{equation}
 \bar{\lambda}_d=An_0^{-3/2}\bigg(1+\frac{\Delta n}{n_0}\frac{y_f^2}{w_0^2}\bigg)^{-1/2},
\label{eqn:lambda_d4} 
\end{equation}
where the initial injection location $y_i=0$, $A = (1/\lambda_L^2)(2\pi c)^3(\epsilon_0 m_e/e^2)^{3/2}$ and $w_0^2/(\Delta n/n_0)$ is defined as the channel width, $r_c$, for practicality. Using a Taylor expansion on the expression in parenthesis in Eq. \ref{eqn:lambda_d4}, it is simplified to Eq. \ref{eqn:lambda_d5},
\begin{equation}
 \bar{\lambda}_d=An_0^{-3/2}\bigg(1-\frac{y_f^2}{2r_c^2}\bigg).
\label{eqn:lambda_d5} 
\end{equation}
where the average dephasing length equals its on-axis value when the radial extend of the beam, $y_f$, is zero.

Given that $\lambda_u$ is the wiggler period and $N$ is the number of periods of the wiggler, to prevent dephasing during the length of the wiggler, one can introduce the relation $\bar{\lambda}_d \ge \lambda_u  N$ and substitute $\bar{\lambda}_d=\lambda_u N$ in Eq. \ref{eqn:lambda_d5} to deduce an average plasma density, $n_{p,d}$, for a given channel width, $r_c$, as given in Eq. \ref{eqn:n0}, 
\begin{equation}
n_{p,d} = \bigg(\frac{4\pi c^3}{w_0^2 \lambda_L N}\bigg)^{2/3} \bigg(\frac{\epsilon_0 m_e}{e^2}\bigg)\bigg(1-\frac{y_f^2}{3r_c^2}\bigg)
\label{eqn:n0} 
\end{equation}
The expression will help determine the plasma density to achieve a certain dephasing length in a channel and  entails two terms. The first term is an on-axis plasma density which can be deduced from Eq. \ref{eqn:lambda_d2} for a $r=0$ and the second term is that scales with the radial extend of the beam and channel width. 

The other factor effecting the dephasing length is the longitudinal velocity reduction of the witness particles due to their sinusoidal trajectories which is approximately equal to $\langle v_z\rangle/c\simeq 1 - (\langle p_\perp^2\rangle+m^2c^2)/2p_z^2$. However, this effect is not considered for the results presented in next section. 

\section{Simulations and discussions}
\label{numerical}
A plasma wiggler and its interaction with a probe beam are simulated using the particle-in-cell code, EPOCH \cite{EPOCH}.  
A 200$\,\mu$m long simulation window is defined with $\pm$70$\,\mu$m transverse acceptance that moves at the 99.9$\%$ of the speed of light. 

The domain is set up with 10 cells per laser wavelength in the longitudinal, and per plasma skin depth for transverse directions, allowing the smallest features in those directions to be resolved. Electrons in the plasma are represented with $5$ macroparticles per cell assuming an immobile neutralizing background. All boundaries are absorbing for particles and fields. Fields are calculated with a $2^\text{nd}$-order Yee solver while currents are smoothed using a 3-point nearest-neighbour low-pass filter in each direction \cite{Buneman1993}. QED photon emission by particles undergoing sinusoidal trajectories is calculated using an optical-depth-like model through local electromagnetic field strengths at macroparticle positions and incorporating radiation reaction and energy straggling effects \cite{Duclous2010}. The wavelengths of classical fields represented by such photons are orders of magnitude too small to be resolved by the grid and are not incorporated into the macroscopic fields of the simulation.

The physical characteristics of the laser, plasma and probe beam are given in the following subsection. 

\subsection{Initial simulation settings} 
\begin{figure}[htb!] 
\includegraphics[width=0.45\textwidth] {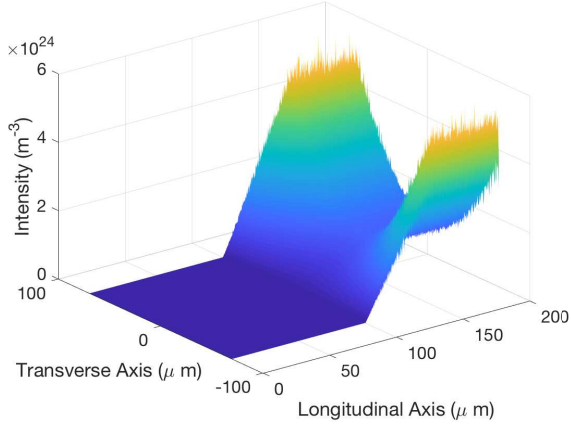} 
\caption{The initial plasma density configuration including a zero density and up-ramp region prior to parabolic distribution.}
\label{fig:density_profile}
\end{figure}

\paragraph{Plasma} A plasma profile comprising a vacuum section followed by a 10$\,\mu$m up-ramp section prior to a parabolic distribution was implemented. The zero density section is provided for initialisation of a laser pulse by ensuring the consistency of the Maxwell's equations. The plasma density profile for the first 200$\,\mu$m of the simulation is presented in Fig. \ref{fig:density_profile}. 

\paragraph{Laser} A custom defined laser pulse is used in EPOCH defining 2D Gaussian beam profiles. Leading order corrections on longitudinal electric field for the diffraction angle expansion as well as for short pulse duration (first derivative of pulse envelope in longitudinal field) are implemented. The model featured control on parameters for focal spot position and temporal pulse centre as well as longitudinal magnetic fields for out-of-plane laser polarisation.

\begin{figure}[htb!] 
\centering
\captionsetup{justification=justified}
\includegraphics[width=0.45\textwidth] {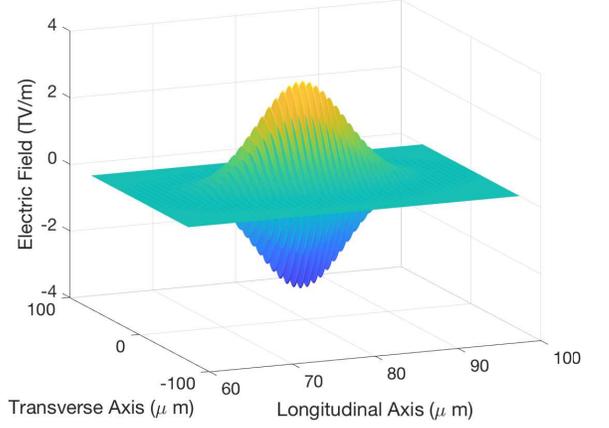} \\
\caption{Initial laser pulse profile for $a_0=0.8$, $\lambda_L=800\,$nm and $w_0=30\,\mu$m.}
\label{fig:laserdefinition}
\end{figure}

The longitudinal pulse profiles are given as,
\begin{equation}
E_{\textrm{pulse}} = - E_0 e^{-(x-x_{t_0})^2/2(\sigma_t c)^2}, 
\end{equation}
\begin{equation}
B_{\textrm{pulse}} = - E_0  e^{-(x-(x_{t_0}-\delta))^2/2(\sigma_t c)^2} /c,
\end{equation}
where $E_0$ is the peak electric field of the laser pulse, $x_{t_0}$ is the spatial position of the temporal pulse centre at $t=0$, $\sigma_t$ is the pulse duration and $\delta$ is the offset for the magnetic field. The initial distribution of the Gaussian laser field can be seen in Fig. \ref{fig:laserdefinition} with a peak electric field of 3.2$\,$TV/m. 

\paragraph{Probe beam} A hypothetical 10$\,$pC electron probe beam with a $\gamma=$1000, a matched spot size of $\sim$1$\,\mu$m and rms length of 1$\,\mu$m. The injection phase is determined so that no accelerating field acts on the probe beam while the focusing field is larger than zero and at its maximum value. The transverse ($E_r$ and $B_{\theta}$) and longitudinal wakefields ($E_x$), that are related to each other through Panofsky-Wenzel theorem, are presented in Eq. \ref{eqn:fields} \cite{Esarey}.  
%
\begin{equation}
\begin{aligned}
E_r - B_{\theta} \sim \left(\frac{4r}{k_p w_0^2}\right)\exp{\left(-\frac{2r^2}{w_0^2}\right)}\sin(k_p\xi) \\
E_x \sim \exp{\left(-\frac{2r^2}{w_0^2}\right)} \cos(k_p\xi), 
\end{aligned}
\label{eqn:fields}
\end{equation}
where $\xi = x-ct$. The conditions, $E_x=0$ and $E_r-B_{\theta} > 0$ can be satisfied simultaneously when $k_p\xi = 3\pi/2$. According to this, the first possible location at $\lambda_p/4$ behind the head of the laser generally is located inside the laser therefore is an undesirable injection spot. Therefore, the next possible location $5\lambda_p/4$ behind the head of the laser is used for the results reported in this paper.  
%
	
\begin{figure}[htb!] 
\centering
\captionsetup{justification=justified}
\includegraphics[width=0.45\textwidth] {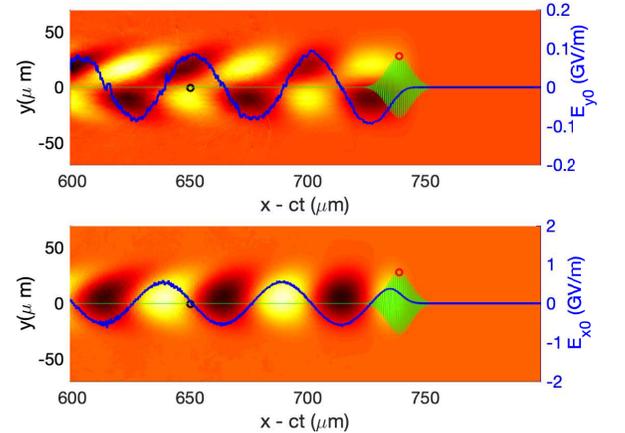} 
\caption{An example case depicting an optimum injection point with $E_x\approx0$ and $E_y>0$. The colour map presents the transverse (top) and longitudinal (bottom) field distributions. The injection location is shown with black circles with respect to the centre of the laser (red circle). Blue line-outs show the on-axis field values.}
\label{fig:injection_locations_sim}
\end{figure}

It is non-trivial to exactly determine the start of the wakefields with respect to the laser pulse front. Therefore, the probe beam injection location is manually studied the field patterns and determined as approximately 89.5$\,\mu$m behind with respect to the centre of the laser. The field distribution experienced by the probe beam is presented in Fig. \ref{fig:injection_locations_sim}. 

\subsection{Probe beam trajectory and radiated power} 
Three merits for initial optimisation of the laser-plasma wiggler prior to injection of a probe beam are effective magnetic field, transverse offset of laser injection point and dephasing length. 

\begin{figure}[htb!] 
\centering
\captionsetup{justification=justified}
\includegraphics[width=0.45\textwidth] {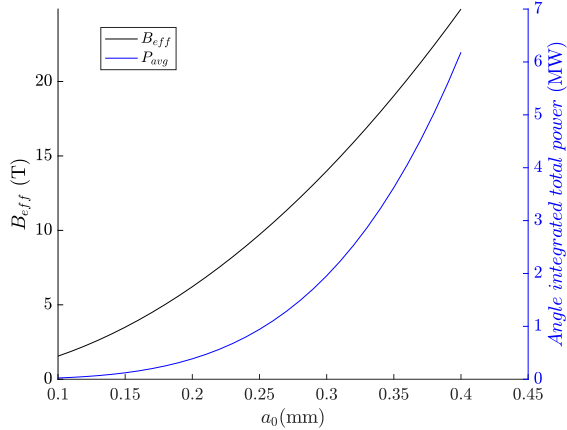} 
\caption{Induced effective magnetic field provided by the plasma wiggler (black curve) and average emitted radiation by the probe beam (blue curve) as a function of $a_0$. The curves are calculated for $w_0=10\,\mu$m and $x_{ci}=6\,\mu$m.}
\label{fig:B_P}
\end{figure}

The effective magnetic field has a strong dependence on the normalised vector field of the laser, injection offset and radius of the laser. The power radiated by the probe beam performing lateral acceleration under this field is given in Eq. \ref{eqn:Pavg}
\begin{equation}
P_{\mathrm{avg}}[\mathrm{W}] = 6.336 E^2\mathrm{[GeV]} B_{\mathrm{eff}}^2\mathrm{[kG]}I_b L_u,
\label{eqn:Pavg}
\end{equation}
where $I_b$ is the peak beam current and $L_u$ is the length of the radiator (wiggler) which is taken equal to the dephasing length for each given wiggler solution in this work.  Figure \ref{fig:B_P} shows that an order of magnitude larger effective magnetic field, compared to current ILC radiators is theoretically possible at around $a_0\approx0.36$, with two orders of magnitude larger radiated average power (4\,MW). 
\begin{figure}[htb!] 
\centering
\subfloat[]{\includegraphics[width=0.45\textwidth]{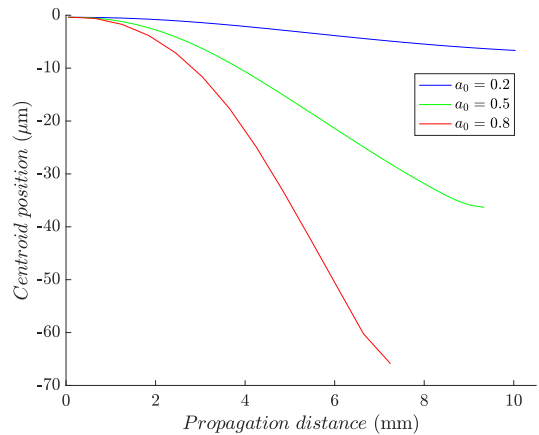}} 
\caption{Beam trajectory following injection at optimum phase for increasing $a_0$.}
\label{fig:ejection1}
\end{figure}

\begin{figure}[htb!] 
\centering
\subfloat[]{\includegraphics[width=0.4\textwidth]{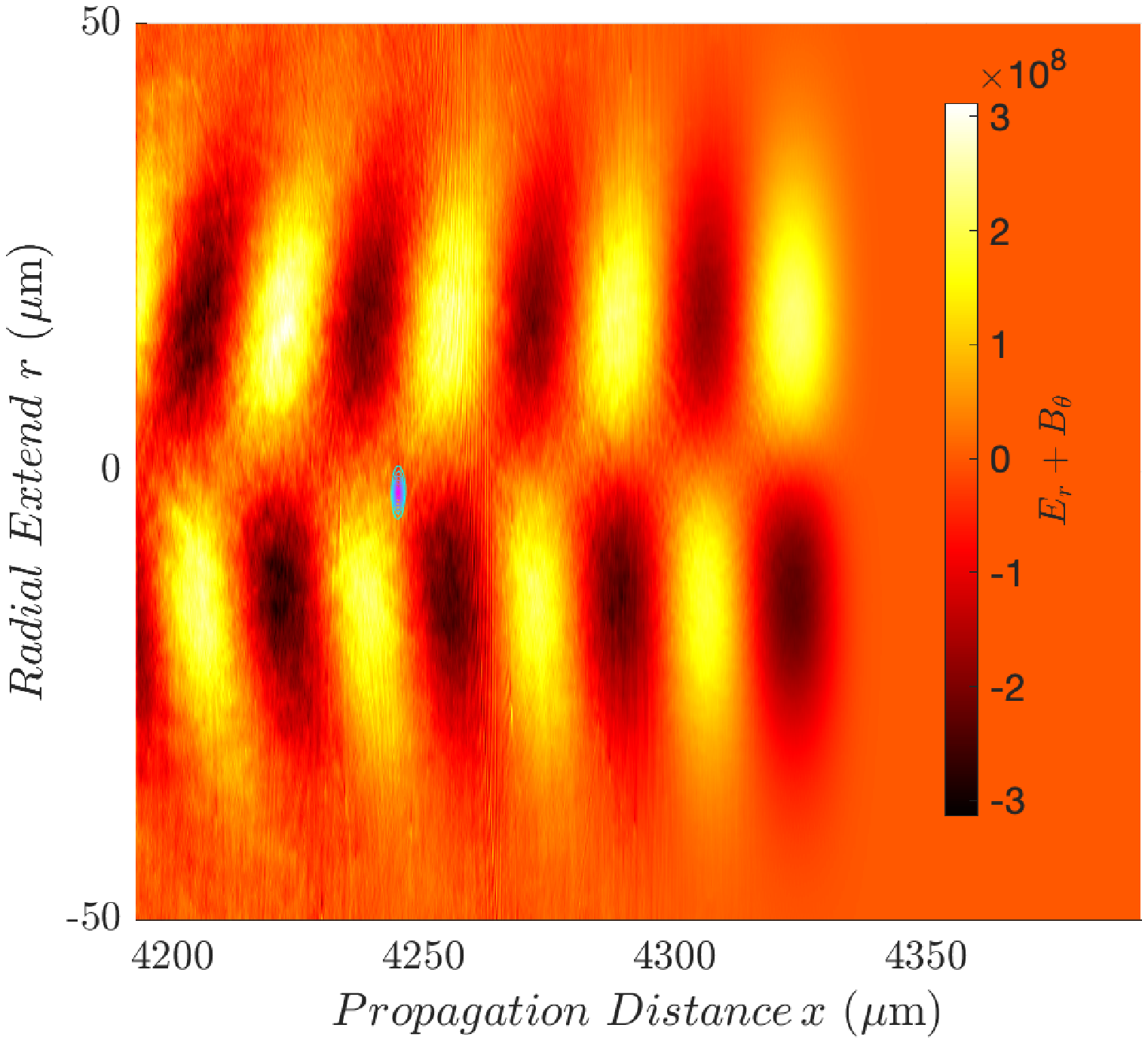}}\\ \vspace{-1em}
\subfloat[]{\includegraphics[width=0.4\textwidth]{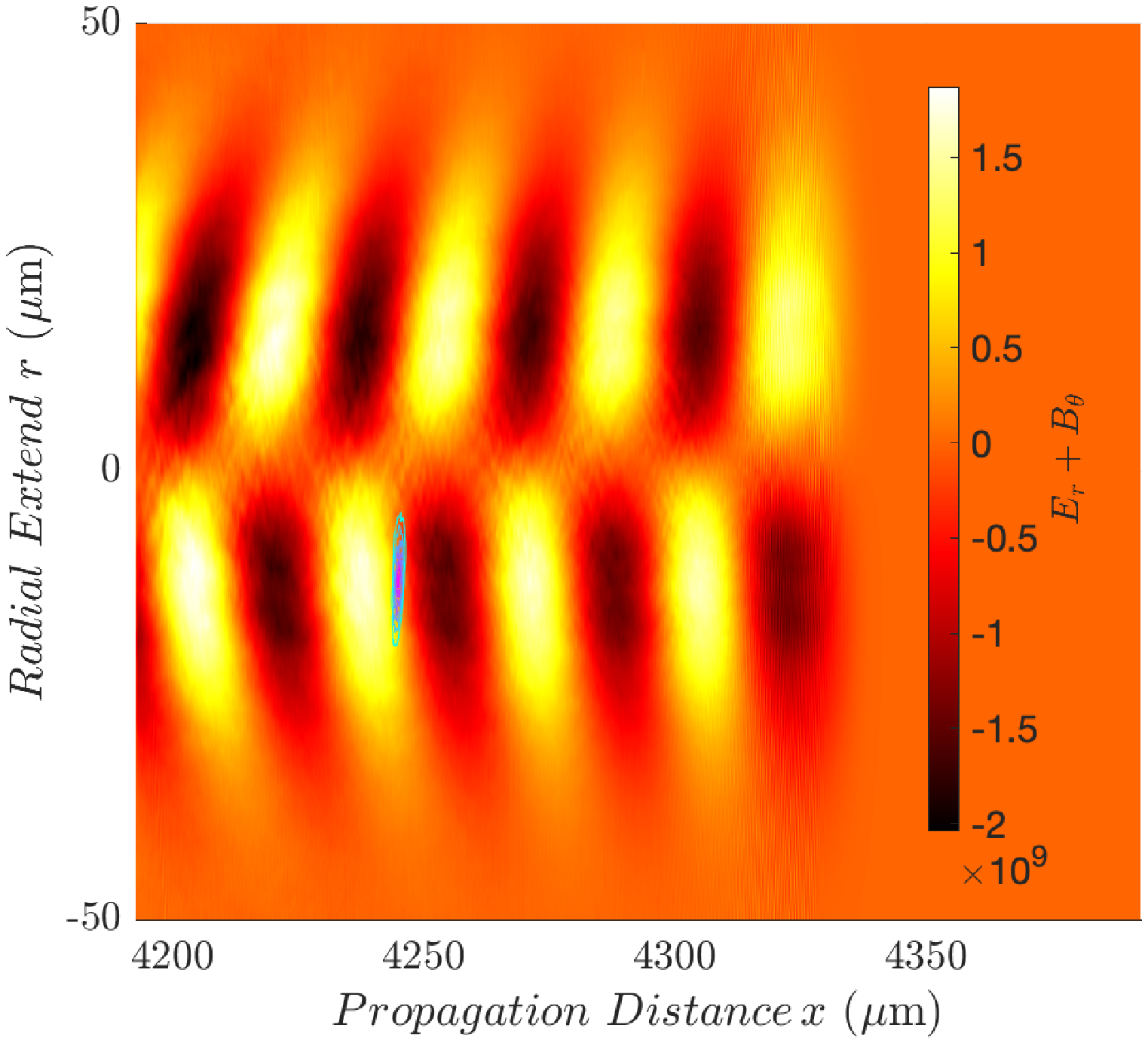}}\\ \vspace{-1em} 
\subfloat[]{\includegraphics[width=0.4\textwidth]{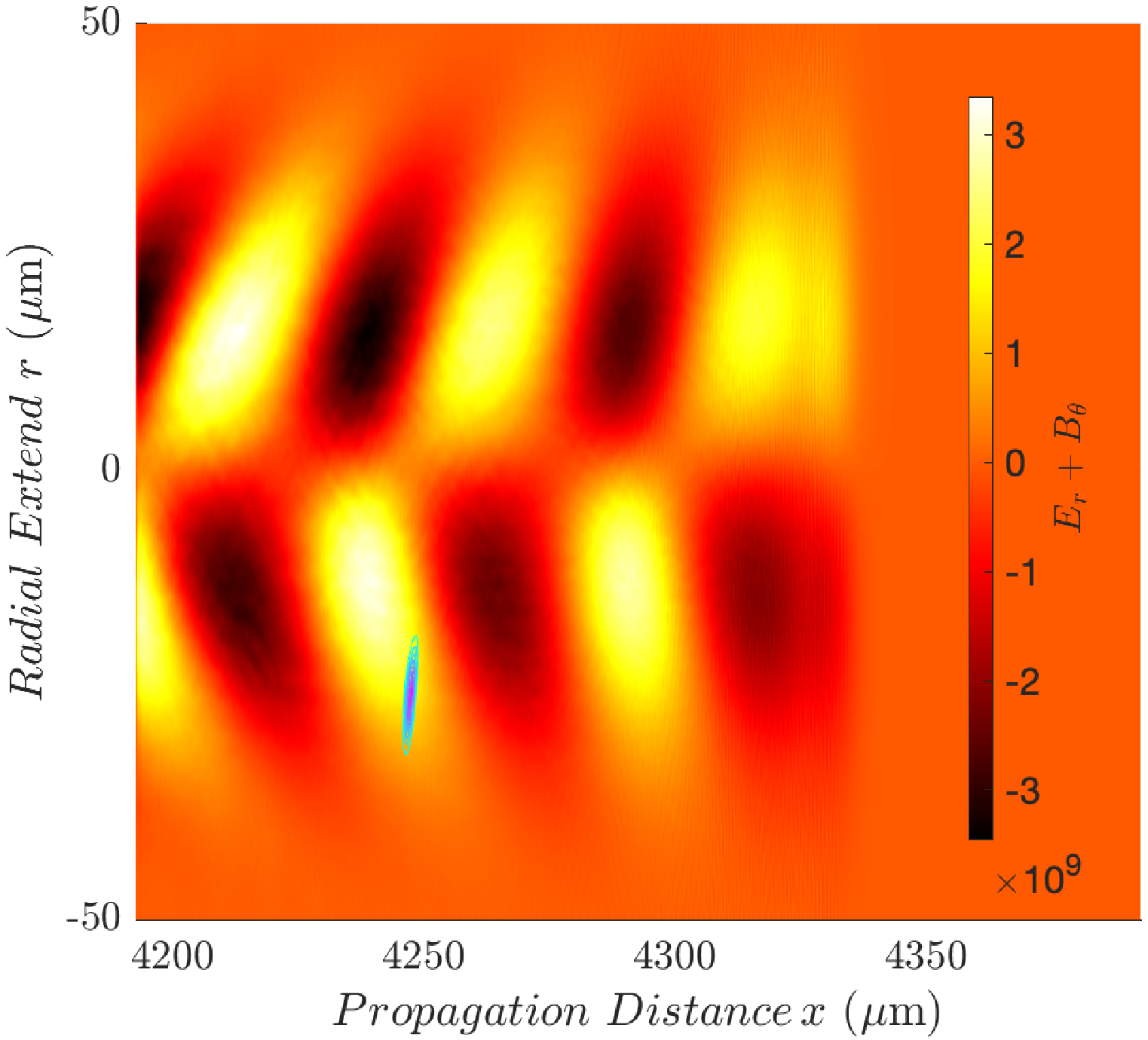}} 
\caption{The ejection of probe beam out of the channel with increasing $a_0$ due to strong transverse focusing field, $eE_y+cB_z$, driven in plasma for (a) $a_0=0.2$, (b) $a_0=0.5$ and (c) $a_0=0.8$ (Table \ref{tbl:summary_solutions}-(I), (II) and (III), respectively). All three cases are for an $x_{ci}=5\,\mu $m, $w_0=30\,\mu $m and $n_e = 1\times10^{24}\mathrm{m^{-3}}$.}
\label{fig:ejection2}
\end{figure}

In summary, in a parabolic channel, high effective magnetic fields can be generated by lasers with high $a_0$. However, there is an upper limit to maintain the probe beam trajectory. The probe beam injected into such a channel might be ejected in one transverse direction before it reaches to the next polarity of its sinusoidal trajectory. Therefore, for a plasma wiggler $a_0$ should be a fraction of unity.  Example trajectories for $a_0=\,$0.2, 0.5 and 0.8 are presented in Fig. \ref{fig:ejection1} and the ejection of the probe beam outside the field regions after a few millimeters of propagation for these three cases are shown in Fig. \ref{fig:ejection2}. Nevertheless, in order to generate bending strengths significantly larger than the conventional technology, one might consider smaller laser spot sizes and hence injection offsets, to compensate against the limit on $a_0$.

\begin{figure*}[htb!] 
\centering
\subfloat[]{\includegraphics[width=0.4\textwidth]{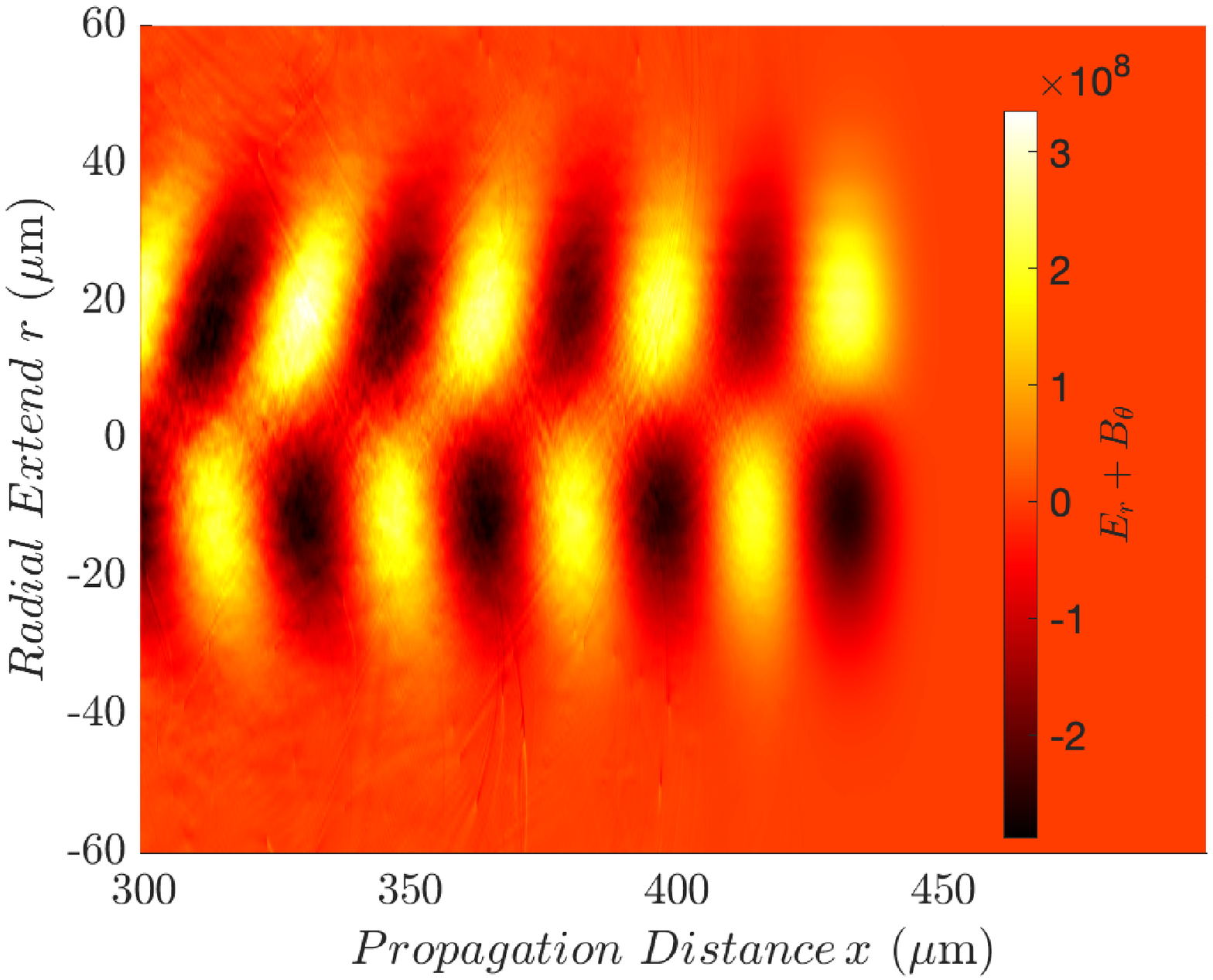}} \vspace{-1em}
\subfloat[]{\includegraphics[width=0.4\textwidth]{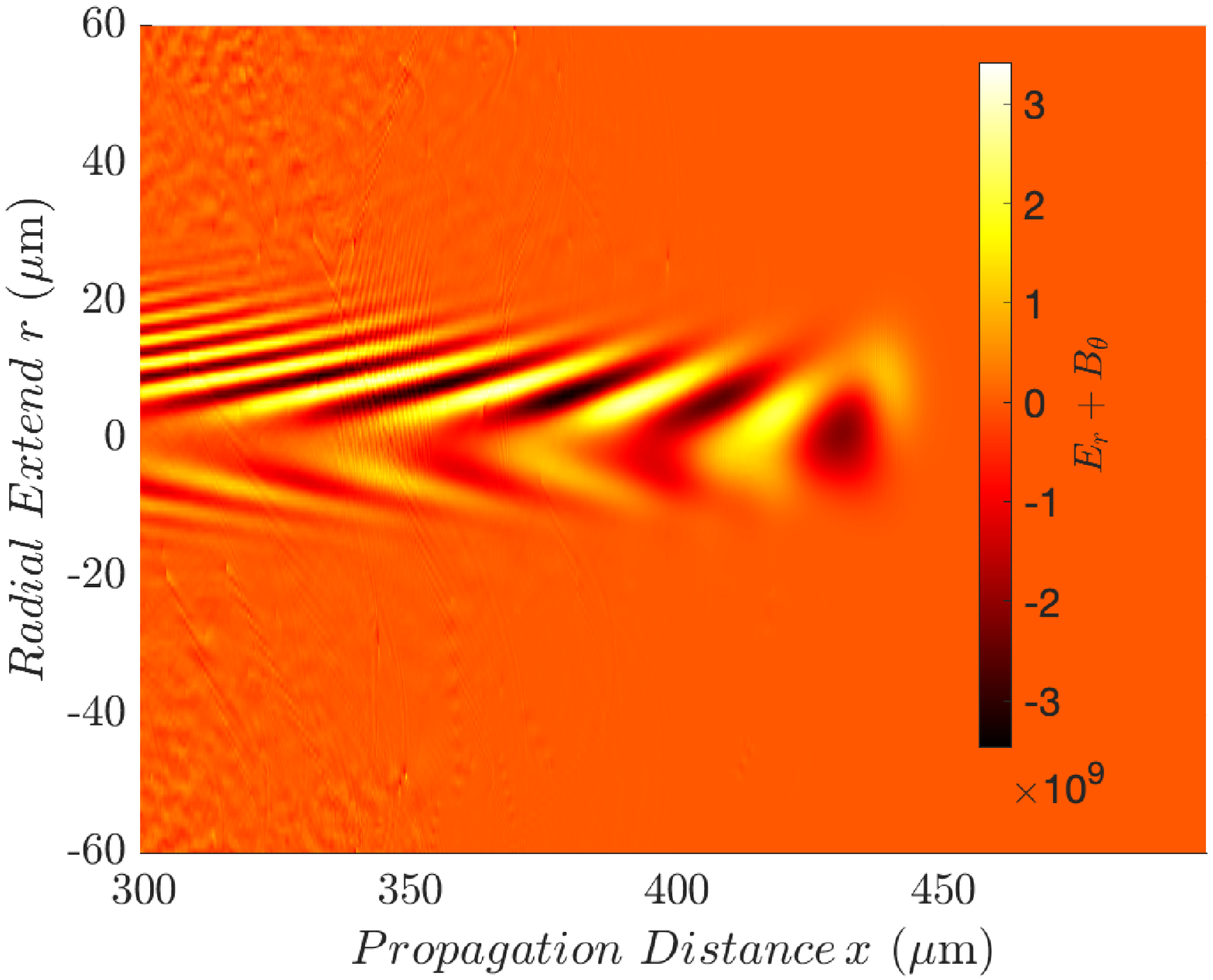}}\\ 
\subfloat[]{\includegraphics[width=0.4\textwidth]{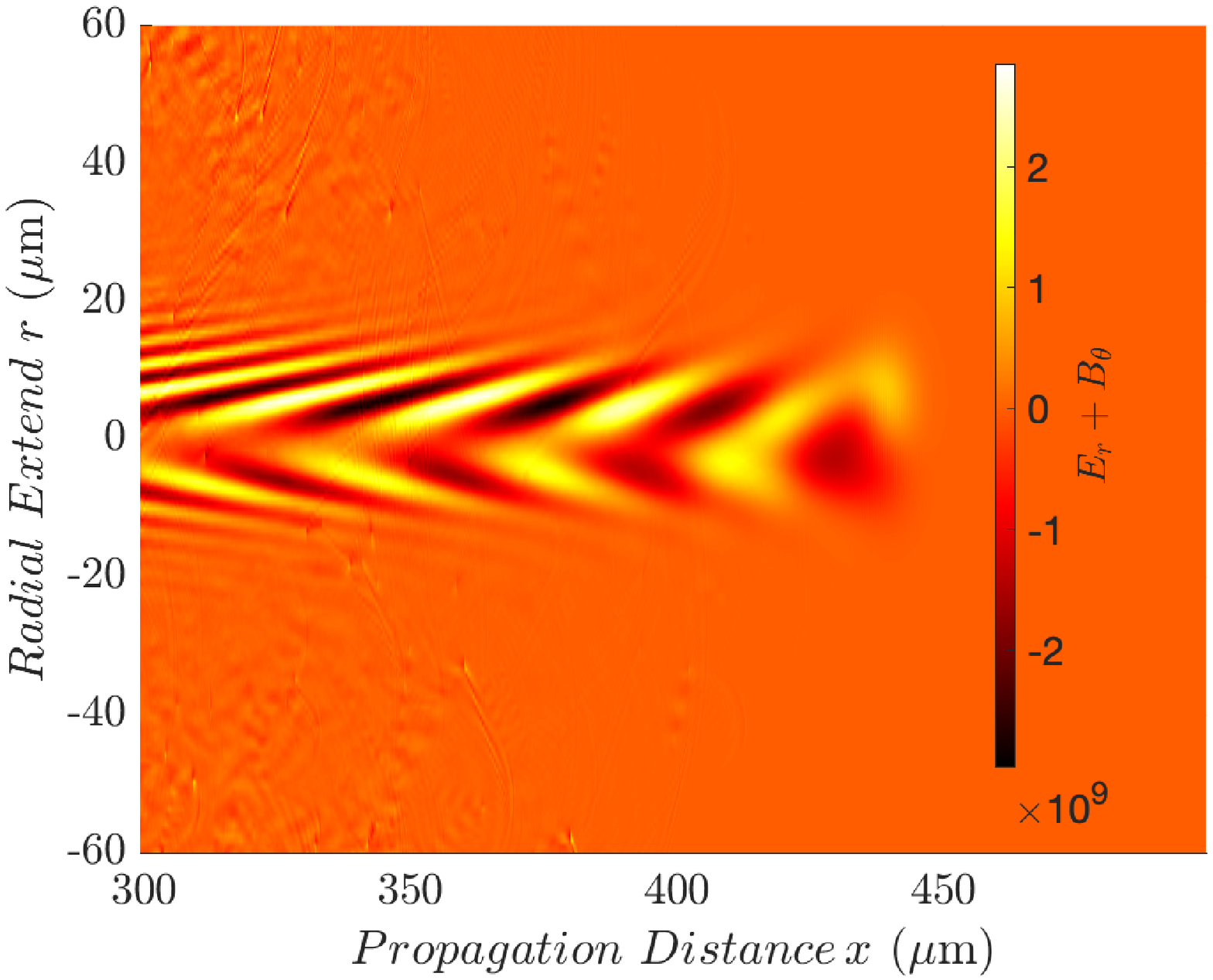}} \vspace{-1em}  
\subfloat[]{\includegraphics[width=0.4\textwidth]{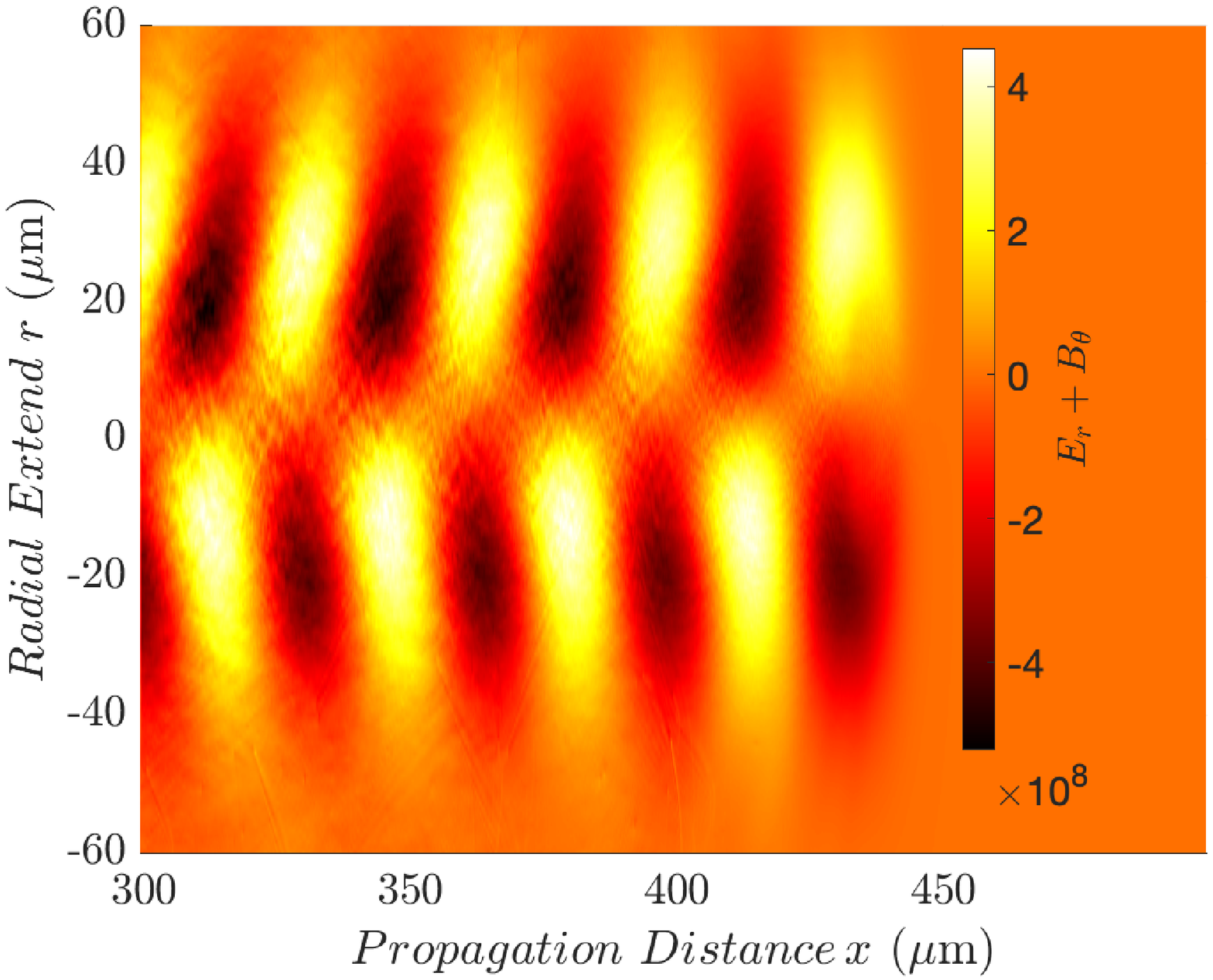}}

\caption{The radial phase of transverse fields for four different wiggler solutions providing a) $x_{ci}/w_0 = 0.2$, $x_{ci}/(c/\omega_p) = 0.94$, b) $x_{ci}/w_0 = 0.6$, $x_{ci}/(c/\omega_p) = 0.94$, c) $x_{ci}/w_0 = 0.3$, $x_{ci}/(c/\omega_p) = 0.5$ and d) $x_{ci}/w_0 = 0.13$, $x_{ci}/(c/\omega_p) = 1$ (Table \ref{tbl:summary_solutions}-(I), (IV), (V), (VII), respectively).}
\label{fig:fieldpatterns}
\end{figure*}

\begin{table}[hbt!] 
   \centering
   \small
   \caption{Analytical solutions for plasma wigglers yielding different radiation power, $P_{av}$, and dephasing length, $\lambda_d$.}
   \resizebox{8.5cm}{!}{ 
   \begin{tabular}{cccccccccc}
    \hline
	Solution & $a_0$	& $w_0\,$  & 	$x_{ci}\,$  & $n_e$	& $a_u$\hspace{1em}	& $\lambda_u\,$ & $B_0\,$ &  $\lambda_d$ & $P_{av}$  \\
	\hspace{1em} & \hspace{1em} & ($\mu$m) & ($\mu$m) & ($\DS{\mathrm{m}}^{-3}$) & \hspace{1em}& (mm) & (T) & (cm) & (kW)\\
	\hline
    \hline

   (I)   & 0.2 & 30 & 5 & $1\times10^{24}$ & 1.2 & 22 & 0.6 & 3 & 2 \\  
   (II)  & 0.5 & 30 & 5 & $1\times10^{24}$ & 7.5 & 22 & 3.6 & 3 & 75 \\  
   (III) & 0.8 & 30 & 5 & $1\times10^{24}$ & 19 & 22 & 9.2 & 3 & 500 \\  

   (IV)  & 0.3 & 10 & 3 & $0.7\times10^{24}$ & 1.6 & 2.5 & 7 & 5 & 500 \\ 
   (V)   & 0.3 & 10 & 6 & $0.7\times10^{24}$ & 3.2 & 2.5 & 14 & 5 & 2000 \\ 
   (VI)  & 0.36 & 10 & 6 & $0.7\times10^{24}$ & 4.6 & 2.5 & 20 & 5 & 4000 \\ 
   (VII)  & 0.3 & 40 & 5.32 & $1\times10^{24}$ & 2.8 & 39 & 0.9 & 3 & 3.5 \\  
    \hline
    \hline
    \end{tabular}}
   \label{tbl:summary_solutions}
\end{table}

Table \ref{tbl:summary_solutions} summarises a set of numerical solutions for phenomena observed during the numerical optimisation of a plasma wiggler to study witness ejection from the channel and radial phase shift of the wakefields. Although, the effective magnetic field scales with the square of $a_0$ and is proportionate to $x_{ci}$, we showed that $a_0 \sim 1$ causes the ejection of witness particles. Hence, angle integrated average power yield through generation of a large $B_\mathrm{eff}$ is achieved by optimising $x_{ci}$. We observed that a radial phase shift is introduced in the transverse fields if the conditions of $x_{ci}/(c/\omega_p)\approx 1$ and $x_{ci}/w_0 \ll 1$ are not met simultaneously. This is demonstrated in Fig. \ref{fig:fieldpatterns} using a set of cases representing different values of the above mentioned ratios. It is also observed that the ratio of the injection offset and the plasma skin depth has a more dominant effect in this phenomena. 

\begin{figure}[htb!] 
\centering
\captionsetup{justification=justified}
\includegraphics[width=0.45\textwidth] {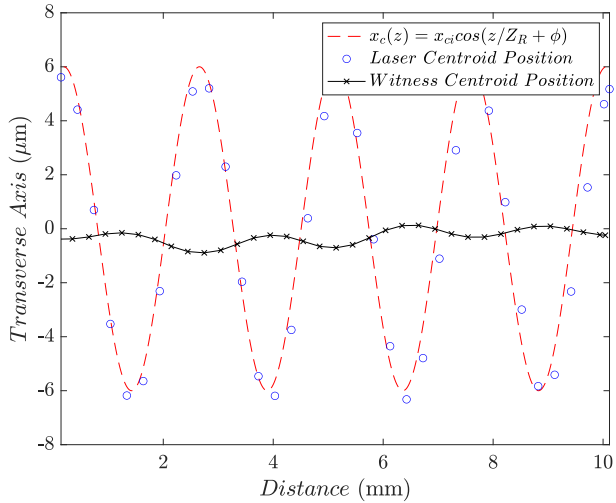} 
\caption{The trajectories of the rms centroids of the laser pulse (blue dots), its theoretical prediction (red dashed line) and the witness for baseline solution given in Table \ref{tbl:summary_solutions}-(VI).}
\label{fig:baseline}
\end{figure}

Counter-intuitively, the radial phase shift seems to accommodate witness propagation at least throughout the dephasing length as demonstrated in Fig. \ref{fig:baseline} for the baseline case given in Table \ref{tbl:summary_solutions}-(VI). In the figure, the perpendicular propagation distance in the plasma is 10\,mm. However, the effective laser propagation is almost 10\,cm due to the sinusoidal trajectory, achieveing the laser dephasing length at about 5\,mm of perpendicular propagation. Figure also shows that the witness centroid follows a sinusoidal trajectory as expected. For this baseline case, the characteristics of the laser, plasma and the probe beam as well as the performance of the resulting plasma wiggler are summarised in  Table \ref{tbl:summary}. We confirmed the feasibility of generating this parameter space in a facility such as Scottish Centre for the Application of Plasma-based Accelerators (SCAPA) \cite{SCAPA} and tests will be planned for future. 

\begin{table}[hbt!] 
   \centering
   \small
   \caption{Characteristics of the initial setup, resulting plasma wiggler and the emitted radiation.}
   \resizebox{7cm}{!}{ 
   \begin{tabular}{lc}
       \hline
       \hline
       \textbf{Laser} & \\
        \hline
       Wavelength (nm), $\lambda_L$ & 800 \\
       Normalised vector potential, $a_0$ & 0.36\\
       Radius at focus($\mu$m), $w_0$ & 10\\
       Pulse length (fs), $\sigma_L$ & 25\\
       Injection offset ($\mu$m) & 6\\
       \textbf{Plasma} & \\
        \hline
        Density ($m^{-3}$), $n_0$ & 7$\times$10$^{23}$\\
       \textbf{Probe beam} & \\
        \hline   
        Normalised beam energy $\gamma$,  & 1000\\
        Density ($\textrm{m}^{-3}$), $n_e$ & $n_0$/1000\\
        Charge (pC), Q & 10\\
        Matched radius ($\mu$m), $\sigma_y$ & 0.6\\
        Bunch length ($\mu$m), $\sigma_x$ & 1\\
        Dephasing length (cm) & 5\\
        \textbf{Plasma wiggler} & \\
        \hline
        Period (mm), $\lambda_u$ & 2.5\\
        Strength parameter, $a_u$ & 4.6\\
        Effective field (T), $B_{\mathrm{eff}}$  & 20\\
        \textbf{Radiation} & \\
        \hline
        Wavelength (nm), $\lambda_{\gamma}$ & 14\\
        Angle integrated average power (MW) & 4 \\
        \hline     
        \hline   
    \end{tabular}}
   \label{tbl:summary}
\end{table}

In this study, we concentrated on numerically demonstrating the feasibility of a plasma wiggler that is predicted by the theoretical optimisation. We demonstrated a baseline setting achieving an order of magnitude larger effective magnetic field (20\,T) and two orders of magnitude larger average radiation power (4\,MW) in a single pass compared to those aimed with a conventional wiggler. 

The phase space cooling occurs after tens of hundreds of passages through the radiator. Hence observation of cooling directly from particle-in-cell simulations is prohibitively computing-heavy with the current capabilities and will be feasible in parallel to studies in exascale computing such as the WARPX project \cite{WARPX}. Therefore, in the next section, we will analytically demonstrate the impact of the designed baseline plasma wiggler on the cooling performance of a linear collider damping ring. 

\section{Transverse Emittance Damping}
 \label{damping}
 
An electron beam circulating in a damping ring losses energy at each turn due to the synchrotron radiation generated from its movement around the circular orbit of the ring as well as the radiation generated from lateral acceleration in the wiggler magnet. The beam emittance decreases exponentially as given in Eq. \ref{eqn:damping},
\begin{equation}
\varepsilon(t) = (\varepsilon_{\textrm{inj}}-\varepsilon_{\textrm{equ}})e^{-2t/\tau} + \varepsilon_{\textrm{equ}},
\label{eqn:damping}
\end{equation}
from its value during injection into damping ring, $\varepsilon_{\textrm{inj}}$, to an equilibrium value $\varepsilon_{\textrm{equ}}$ where $\tau$ is the damping time.

The emission of synchrotron radiation is a statistical process involving quantum fluctuation of beam parameters. An equilibrium emittance is reached when the quantum excitation is equal to damping  \cite{handbook}. 

\subsection{Contribution from wigglers}
The energy loss in wigglers constitutes nearly 95$\%$ of the total energy loss. This energy loss of a beam during its each passage through a wiggler is given in Eq.\ref{eqn:U0} 
\begin{equation}
U_0 = \frac{C_{\gamma}}{2\pi} E^4 I_{2w}
\label{eqn:U0}
\end{equation}
which is proportional to the forth power of its energy $E$ and second radiation integral for a wiggler $I_{2w}$. This integral can be written as in Eq.\ref{eqn:int2} 
%

\begin{align}
I_{2w} = \int_0^{L_u} \frac{1}{\rho^2}ds
= \frac{1}{(B\rho)^2} \int_0^{L_w} B_w^2ds  = \frac{1}{(B\rho)^2} \frac{B_w^2L_w}{2}.
\label{eqn:int2}
\end{align}
for a wiggler with an effective length of $L_w$ and a radius of curvature $\rho$ for the deflected trajectory. To evaluate $I_{2w}$, beam rigidity $(B\rho=p/e)$ is used to simplify the integral. 

From Eq.\ref{eqn:int2}, it is straightforward to see that a wiggler contributes to the energy loss per turn in a damping ring proportionately to the square of its magnetic field and the effective length, i.e., 
\begin{equation}
U_0 \propto B_w^2 L_w.
\end{equation}

Furthermore, the damping time, which is the time needed for the emittance to decrease by 1/e of its injected value, is given as,
\begin{equation}
\tau  = 2E_0 \frac{T_0}{U_0},
\end{equation}
where $E_0$ is the injection energy of the electrons into the damping ring and $T_0$ is the revolution period of the electrons in the ring. This implies that an order of magnitude larger field enables the same amount of damping with a 100 times shorter wiggler. Alternatively, for the same radiative length, a 100 times faster damping might be achieved with a wiggler providing an order of magnitude larger wiggler strength.

\subsection{Cooling scenarios for an ILC-like machine}

\begin{figure}[htb!] 
\centering
\captionsetup{justification=justified}
\includegraphics[width=0.46\textwidth] {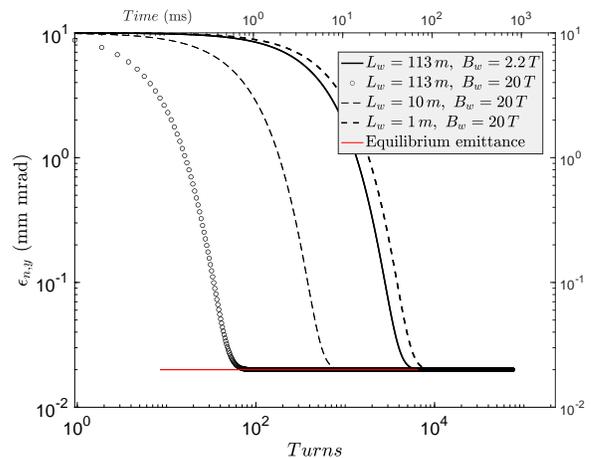}
\caption{Performance of different cooling scenarios using plasma wigglers for ILC damping ring in comparison to baseline with superferric wigglers.}
\label{fig:damping}
\end{figure}

The impact of a plasma wiggler on the damping process was studied using parameters similar to the baseline for ILC damping ring to demonstrate its potential for future colliders. To study this, an injection emittance of 6$\,$mm$\,$mrad was considered to be damped down to 20$\,$nm ($\mu$m$\,$mrad). The ILC-like baseline case consists of a total radiative length of 113$\,$m including 54 wigglers with 2.16$\,$T (as defined for electrons at ILC in the technical design report). The evolution of emittance is studied as described in Section \ref{damping} and presented in Fig. \ref{fig:damping} for different scenarios. For ILC damping rings, the equilibrium emittance should be reached below 200$\,$ms, between machine pulses, which is shown in solid black curve in Fig. \ref{fig:damping}.

The first scenario proposes to replace the superferric wigglers by plasma wigglers with characteristics given in Table \ref{tbl:summary} over the same 113$\,$m radiative section. This is presented with hollow dots and provides the largest reduction in damping time below 1\,ms. The second scenario employs the advantage in significantly increased bending strength and proposes to decrease the radiative section length an order of magnitude down to 10$\,$m, that achieves damping within about 5\,ms (dashed light curve). Eventually, the third scenario reduces this down to a one meter-long radiative section (dashed bold curve) for a larger damping time than the conventional ILC option. Nevertheless, it still provides an 80\,ms damping time which is adequate for the ILC machine cycle. 

A damping ring design with a higher ring energy provides shorter damping times. However the natural emittance scales with the cube of the energy deeming a lower energy ring favourable. Keeping in mind that the collective effects are more severe at low energies, the plasma wiggler offers a solution where at least two orders of magnitude shorter damping times are possible while keeping energy low enough to ensure a minimal natural emittance for the ring. Therefore a next generation damping ring equipped with plasma wigglers rather than conventional counterparts provides an exciting scenario for the future linear collider design studies.

\subsection{Multi-stage single-pass cooling}

The circumference, hence the cost of a damping ring increases with the number of bunches per linac pulse. The ILC design envisages bunches to be injected into a ring in a compressed time structure to limit its circumference. After cooling, these bunches should be decompressed by using state-of-the-art dipole magnets (kickers) during extraction to provide the desired bunch spacing for the experiments. Previous studies showed that, in a damping ring, certain instabilities are introduced during injection/extraction \cite{Apsimon2014} and circulation until a design orbit is established before cooling. 

Furthermore, due to the quantum emission, there is an equilibrium energy spread of the beam that is established in the ring. Because the ring is dispersive, this energy spread results in an equilibrium bunch length. This equilibrium bunch length produced in conventional damping rings (9 mm for compressor entrance) is much longer than the typical values for operating conditions of a plasma-based accelerator. Reducing the bunch length is difficult, and results in intra-beam and Touschek scattering (due to high current), which are sources of unacceptable emittance growth. Therefore one must use bunch compression after a damping ring. For example, ILC compresses about a factor of ~30 after the ring. For a plasma accelerator requiring bunch lengths a fraction of the plasma wavelength, e.g., as small as $\sim$10 microns, this would require bunch compression factors of order $\sim$1000. It is difficult to imagine any bunch compression system that could achieve this while still preserving the beam emittance. 

As such, a scheme eliminating the use of a ring altogether would revolutionise the cooling process for a collider based on plasma technology. This future work will require extensive computing-heavy numerical work to explore the beam dynamics of a single-pass linear damping section with adjacent plasma stages for radiation and acceleration. Staging is already a hot topic in the field, providing the community with many challanges to tackle \cite{stage1, stage2, stage3, stage4, stage5}. 

\section{Conclusions}
An innovative concept of using a plasma wiggler for beam damping was explored.  Initial results demonstrate an order of magnitude larger effective magnetic field achievable with realistic laser and plasma parameters. This has direct implication of several scenarios compromising between orders of magnitude faster damping or smaller footprint. A comparison to the ILC damping ring design, as a conventional example, was presented demonstrating, for extreme scenarios, damping within $<$1\,ms or alternatively reduction of footprint of radiative section down to 1\,m from 113\,m. 

Using plasma wigglers for emittance damping brings an exciting prospect for future linear  colliders especially based on plasma technology. This may include muon colliders as well, where cooling time might compete against decay time. This opens up new avenues for exploration of the implementation at large-scale, numerically, where reduced models or extensive high performance computing resources will be required as well as studies regarding the multi-staging. 

The reduction in damping time provided by a plasma wiggler can compensate against the need for large energy in a conventional set-up, allowing to limit the natural emittance in the ring. That being said, implications on feasibility of the concept regarding instabilities in a damping ring at low energies will be explored. 
\\
\begin{acknowledgments}
This work is supported by the European Union's Horizon 2020 research and innovation programme and U.S. NSF grant 1804463 and AFOSR grant FA9550-19-1-0072. Computing resources provided by STFC Scientific Computing Department's SCARF cluster. Authors would like to thank to Jon Roddom and Derek Ross for their endless support on implementation of EPOCH on SCARF clusters. We would like to thank to EPOCH community support, especially, Keith Bennett, Thomas Goffrey, Heather Ratcliffe and Christopher Brady. Authors also would like to thank to Dr Carl B. Schroder (BELLA Centre, Berkeley Lab), Dr Hywel Owen (The University of Manchester) and Mr Volodymyr Rodin (The University of Liverpool) for insightful discussions on the requirements of a damping system for beams generated by plasma-based accelerators and storage rings. 
\end{acknowledgments}

\bibliography{apssamp}

\end{document}